# Interaction semantics

# – and its implications for an interaction oriented architecture of IoT-type applications.

## 2017-01-29 (revision 1 of 2017-05-16)

Author: Johannes Reich, SAP SE

## Summary

Several synergistic trends, subsumed under the phrase "Internet of things (IoT)" massively drive the increasing importance of networking applications. In the past, the exponential growth of the Internet was mainly due to semantically agnostic transport protocols. In the future it is to be expected that – because of the increasing autonomy of technical systems – it becomes necessary to better understand the nature of the semantics of these interaction networks to create appropriate networking applications. Appropriate means that the architecture of these applications allows to minimize the effort to adapt these applications to the permanently changing interaction networks.

The proposed interaction oriented architecture is based on a reference model of interaction semantics. It provides guiding principles on how to design networking applications.

The reference model of interaction semantics provides:

- A unifying description of the things in the physical, the information and the human world.
- An interaction model that is of direct runtime relevance.
- An understanding for how hierarchical structured components can cooperate loosely coupled.
- A concept to determine how much semantics has to be common to enable components of different semantic domains to cooperate loosely coupled.
- A data type model.

The software reference architecture provides

- A definition of software layers
- Means to express vertical interactions, that is interactions which demarcate a software layer.
- Means to express horizontal interactions, that is, between processes in the same software layer.
- A definition of a component and how to distinguish it from other entities like systems or objects.
- A model how to separate reusable from non-reusable parts of an application's functionality.

# Contents

## Introduction

With "Internet of Things (IoT)", several synergistic trends in the area of electronically interacting devices are subsumed:

- There is a dramatic increase in the number of electronic devices.
- The interconnectivity of each device increases.
- The interactions with humans and the environment increase.
- The interactiveness of each devices increases.
- The devices become more and more autonomous.

As a result, the already vast network of interconnecting devices will further increase enormously. As network effects are one of the strongest known economic forces, there is no doubt that IoT in this sense will transform the way we live and conduct business in many ways.

In general, these interaction networks are open and change permanently. Their permanent change poses a major (software) engineering problem. Although single classes of interactions often remain remarkably stable, the nodes of these interaction networks usually take part in multiple interactions. A simple calculation emphasizes this dynamic: a relatively low change rate of 10% of the interaction classes per year together with an average of 10 different interactions per node, it has to coordinate, would require, roughly estimated, to change on average each node each year approximately once!

Fig 1 shows two cutouts of such networks as examples. Using the same template for both cases, of enterprise applications as well as business partners, underlines the universality of this networking idea. The naming of the nodes of these interaction networks depends on the context. We use "device" if we refer to something we can put our hands on. We use "node" if we refer more to the topological aspects. We use "application" if we refer more to some software engineering aspects. "System" emphasizes the universality, abstracting away the differences between machines and humans. The term "reactive systems" was introduced by David Harel and Amir Pnueli [HaPn1985] to delineate the fact that from an interaction perspective this kind of system "do not compute or perform a function, but is supposed to maintain a certain ongoing relationship, so to speak, with its environment.". The same connotation can be attributed to the term "process".

A more traditional example of a complex network change occurs in a merger of companies, where two or more IT-infrastructures have to be merged. Or, if a company wants to integrate a logistics component in addition to the already existing components for enterprise resource planning and customer relationship management. How much will the old components have to be changed?

Another example could be a telecommunication solution. Here a CRM may manage the customer facing activities, like contract management, for the telecommunication provider. It interacts with its ERP system performing all financial and logistics activities. This is sufficient for sales and service scenarios as long as no other interaction partners needs to be involved. However, in contract creation additional contract information (e.g. technically valid phone numbers) and an "activation date" (on which the service is technically enabled) suggests additional interactions with the technical telephony systems. The overall interactions may get even more complex with requirements that the telco company has to stop the service, once the account balance is zero (prepaid), etc.

So, from a (software) engineering perspective, three essential questions are:

1. How far do these changes vibrate through the whole network of interactions?
2. How much effort has to be invested to change the affected nodes?
3. How can we guarantee scaling of our system design with the size of the networks?

Cutout of a network of enterprise applications

Cutout of a network of business interactions

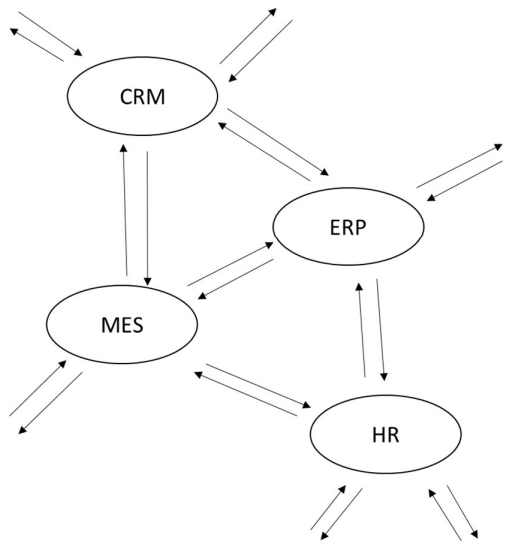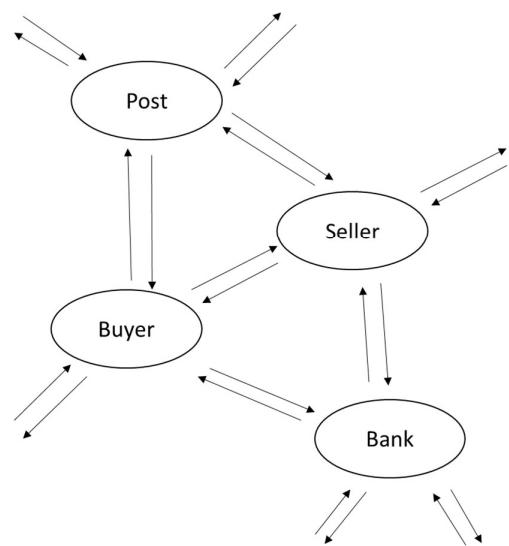

*Fig 1*: *Cutouts of two networks of interactions with the same topological relations. On the left, a network of enterprise applications is shown, dealing with manufacturing: a manufacturing execution system (MES), an enterprise resource planning system (ERP), a customer relationship management system (CRM) and a human resource system (HR). On the right a cutout of a network of business interactions is shown. A buyer interacts with a seller, a bank (for paying) and a post (for delivery).*

In the past, the exponential growth of the internet was mainly due to semantically agnostic transport protocols like HTTP, FTP, SMTP, IMAP, etc. The content was mainly processed by humans, keeping the problem of semantic interoperability within the human interaction sphere. In the future, it is to be expected that the mentioned trends will put a vigorous pressure on rendering semantic interoperability achievable with minimal effort also on a machine level. As will be shown, the Interoperability semantics is tightly related to the processing of the interchanged data. In the IoT, data processing is mainly achieved by software. So, it will be the software-part of the engineering disciplines that will provide the most relevant answers to these questions.

To answer these questions and to provide a solution to the engineering challenge, the nature of the semantics of the network interactions, or, in other words, the problem of interoperability on a semantic level has to be understood. This insight is one of the essential differences to the past.

The big challenge from an engineering perspective is not to create somehow functioning applications in these interaction networks. Such a naïve viewpoint would entail a substantial risk that we will not be able to cope with the speed of change and end up with unsurmountable integration problems – even hindering the connectivity dynamics. Instead, we have to understand how to build applications which are as robust as possible against the changes of the interaction network they are part of. That is, I propose that robustness against change of interactions should be viewed as one key quality for applications in the sense of [Reich2016b]. Here, the interaction oriented architecture wants to provide useful concepts.

## Interaction perspective          ## Process perspective

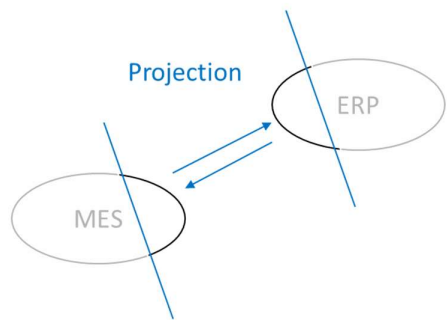
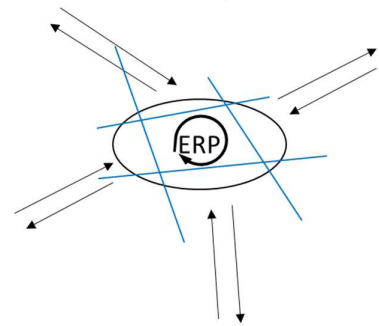

*Fig 2: Two perspectives can be distinguished for describing networks of interacting processes: the interaction perspective, where only a projection of the involved processes, the roles, become directly visible to the other processes, and the process perspective, focusing on the coordination of the different roles within a process.*

The central idea of the interaction oriented architecture is to understand the semantics of these interaction networks as an interaction semantics where every sender expects its sent information to be interpreted in a given receiver context, providing the "meaning" to the exchanged information. This context is the I/O-relation of the receiver as far as it is known to the sender. So, instead of saying that we have to understand the semantics of the transmitted data, we could equivalently say that we have to understand how the I/O-relations of the interacting parties relate to each other.

I call an unintended mismatch between the sender's expected and the receiver's real transition relation a misconception or a misunderstanding, depending on whether the sender or the receiver made the fault. An intentionally precipitated mismatch I call a deception.

The basic criterion for semantic interoperability becomes substitutability: One component is semantically equivalent to another component in an interaction, if it can replace it in this interaction. A necessary condition is that it provides the same transformational behavior, or I/O-relation[1] with respect to the expectations of the sender.

In effect, for interacting processes, we introduce a second perspective beside the traditional system perspective, namely the interaction perspective (see *Fig 2*).

From the traditional system perspective we look at what a system does: it acts! An action in this sense is the transformation of the world by a system. For an engineer it is essential that the I/O-relation of a system defines a function, as by this fact, a system becomes constructible.

Exposing the system's function e.g. by a traditional application programming interface (API) means that the caller has the expectation that if she provides a certain input value, it will get transformed according to the API-semantics into an appropriate output value. This understanding of semantics is in full accordance with the semantics of traditional imperative programs, describing computable functionality.

However, in interaction networks we deal from a software engineering perspective with processes (or "reactive systems") in the sense that in these networks systems receive multiple inputs and provide multiple outputs in a way that their behavior becomes nondeterministic from the point of view of a single interaction partner. As an important consequence, even if each process in the network may work deterministically according to its system function – from the perspective of each interacting partner system, only a part, namely a projection of this function becomes visible (see *Fig 2*). Then the additional perspective makes perfect sense.

---

[1] We restrict ourselves to the relevant transformational I/O-behavior. De facto, quality aspects relating to the system context like performance, security, etc. will also be important for the semantic relations between systems.

From the interaction perspective, we look at what a sender system knows that a receiver system can infer from what it (the sender) just had said. Due to the lack of knowledge of the interaction partners, the execution of a process' system function becomes interpretable only in the sense of an event from the interactors' perspectives. This does not mean that the still necessary knowledge is in some sense fuzzy or vague. At least the rules relating to the I/O-behavior are, although incomplete, still precise[2]. They are documented in interaction protocols, where each participating system is fulfilling a role. Thereby protocols represent the part of common knowledge which has to be shared for mutual understanding as far as it relates to the mutual interactive I/O-behavior. For example, a „customer" has to have the same meaning for an ERP as well as for a CRM system only in so far as it is relevant for achieving their common interaction goal. This is by far less demanding than the requirement to share the complete object model "customer" by both applications.

Now, the important insight is, that in process interactions, where only a projection of the system's function becomes effective, the semantic criterion of substitutability is much less demanding compared to the case where the complete system function is exposed: processes are only "loosely coupled" in the sense of the different sizes of the state space of the acting processes compared to their interactions. If a system exposes its system function completely, both spaces are identical.

If the transition from the process perspective to the interaction perspective is projection, how do we get from the interaction perspective to the process perspective? By coordination: a process can be viewed as a coordination of all its different roles it is involved. Such a coordination can be viewed as a set of rules relating the events of the different roles of a single process. Reuse of roles for different processes by different coordinations is one of the most interesting aspects of the interaction oriented architecture.

One price we have to pay for this semantic loose coupling is the property of interactivity, which comprises two important facts: the interactions of processes will generally be nondeterministic, asynchronous, and stateful. And second, we can tell whether some processes show a sensible behavior only if we know the abilities of its interaction partners.

Abstractly speaking, we gain scalability for our descriptions of interaction networks by sacrificing global knowledge of the networks functionality and limit ourselves to the local knowledge of the (interactive) interactions.

Remarkably, both perspectives can be expressed by different syntactic means: functions can be denoted by traditional APIs, while such interactive interactions, best denoted as "mutual hinting" by events are described by protocols. This is illustrated in *Fig 3*.

---

[2] This in in contrast to natural interactions of humans, where vagueness and a permanent change of interaction levels and rules are possible. I think that such kind of uncertainties requires the application of artificial intelligence techniques and heuristics and goes beyond the presented concepts of an interaction oriented architecture. However, I would expect that a better understanding of the structure of formally describable behavior in interaction networks also improves our sense of application of these other technologies in this domain.

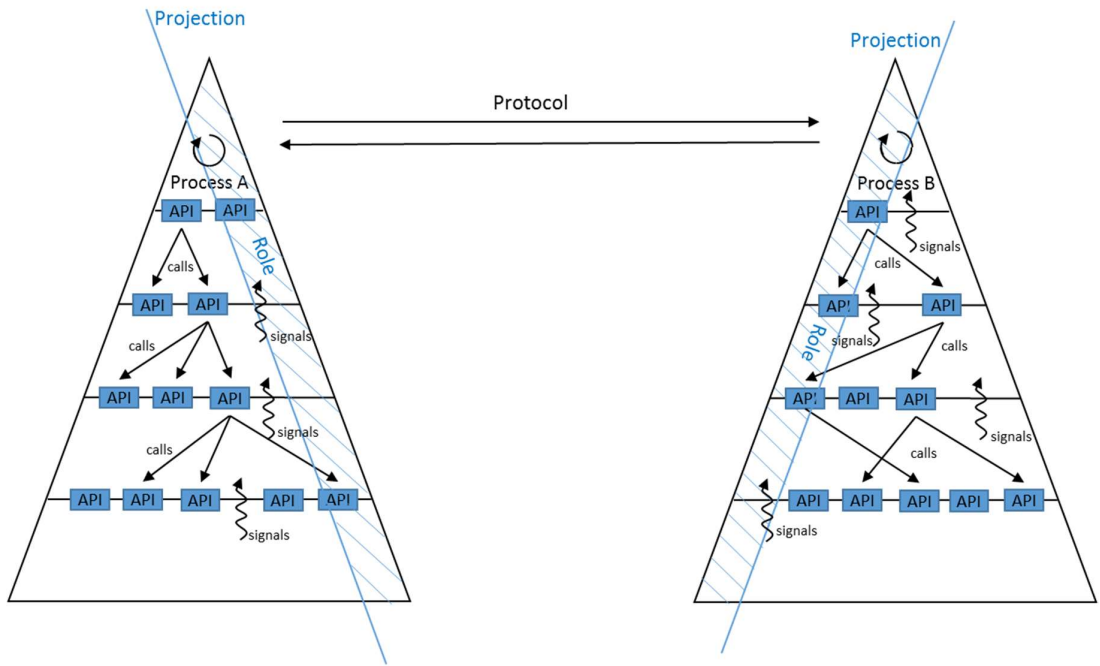

*Fig 3: Schematic representation of the layered application structure, supported by APIs, internal events and protocols. The hierarchical functional composition of functionality is described by APIs. The bottom-up interaction is accomplished by an event semantic. Loosely coupled interactions between two such applications are described by a protocols. By protocols, only a projection of the process' functionality becomes effective from the interaction perspective.*

The software engineering challenge to know how to build applications which are as robust as possible against the changes of the interaction network they are part of can therefore be reformulated: how to architect computational processes – which work deterministically in the sense that they have a unique I/O-relation in each step but generally provide nondeterministic, stateful and asynchronous behavior to each network partner – such that changes in single interactions leads to minimal possible adjustments in as few as possible processes.

In the following, the reference model and the reference architecture are described in more detail to provide guidance to software and system engineers when defining their system architectures.

## The reference model of interaction semantics

The proposed reference model of interaction semantics for IoT software architecture provides:

- A unifying description of the things in the physical, the information and the human world.
- An interaction model that is of direct runtime relevance.
- An understanding for how hierarchical structured components can cooperate loosely coupled.
- A concept to determine how much semantics has to be common to enable components of different semantic domains to cooperate loosely coupled.
- A data type model.

### A unified description of things in the world based on the system notion

The unified description of the things in the physical, the information and the human world is based on the system notion. The system notion is based on a physical terminology. Our physical descriptions of the world rest on time dependent properties, that is functions $f \colon T \to D$ mapping time $T$ onto some value domain $D$. This can be a position, a velocity, a current, a voltage, a density, etc. Different scientific domains have created different terms to denote this class of entities: state variables, state quantities, state functions, process variables, signals, etc.

However, we are not interested in the values of pressure, voltage or current as such, but only in that "something" that can be transferred onto other state quantities. This leads to the information concept as it was formulated by Claude E. Shannon [Shannon1948] and others. As a consequence, I would like to reserve the term "state quantity" for the physical entity, measurable in volts, amperes or bars and use the term "signal" for the same entities viewed through the informatics glasses. I say that a signal can represent a (dimensionless) value at a given time. I call these signal values also "state values" to be more aligned with traditional automata theory terminology.

We have to be aware that the semantics of these values is invariant against renaming. That is, if a signal can represent three values, we can name them "cars", "people" and "planets" but without further context, we do not say something different if we rename the values to "a", "b" and "c" or to "1V", "2V" or "3V". We could therefore even exclusively use our names for numbers. So, it's not the concrete determination of these terms that is relevant for our semantic model but only the fact that a determination is necessary to talk about the signal values – and we have to look very closely where our semantic model relies on the convention of identical names.

Classifying a state value as "information" in the context of transport paves the way to classify it as "data" in the context of processing, as will become clearer in the following. And it also allows to define in a precise sense what we refer to by "semantics of interactions", which integrates the human world of meaning. In effect it shows that these "different worlds" are only different ways to describe the one and only world we are part of (for the formal details, see [Reich2016]).

If the relation between certain signals can be described by a function, then we speak of a system. This definition is in line with current system theory [e.g. LeVa2011, Lunze2012]. It's this signal mapping functionality that is system constitutive, allowing for provable system borders and for a defined composition behavior of systems to super systems.

The structure of $T$, $D$ and $f$ allows to classify systems. The time domain can be discrete or continuous, $D$ can be finite, infinite or can allow only atomic elements versus sets. $f$ can be computable, finite, analytic, allow for causality, etc.

In the following I will assume a discrete time scale, signals representing only single values and a computable functionality with the special property that the mapping between the signals can be reduced to a mapping between signal values at two consecutive points in time. We thereby gain a simple time scale and causality. I usually will mean this function, when I use the term "system function".

This restriction does not restrict us from describing the interactions with systems of other system classes, as these interactions will be described by system projections. It is easily possible to project a continuous time on a discrete scale or an infinite set on a finite one. So, for describing interactions to systems of other classes, even very complex ones like humans, such potentially partial knowledge suffices.

So formally, a system in this sense is characterized by the tuple $(T, succ, Q, I, O, q, in, out, f)$, where $Q$ is a non-empty set of internal signal values and $I$ and $O$ are the possibly empty sets of input and output signal values (also named 'alphabets'). The signals $(q, in, out) \colon T \to Q{\times}I{\times}O$ are said to form a (discrete) system for time step $(t, t' = succ(t))$ if they are aggregated by a function $f: Q \times I \rightarrow Q \times O$ with $f = (f^{int}, f^{ext})$ such that

$$\begin{pmatrix} q(t') \\ out(t') \end{pmatrix} = \begin{pmatrix} f^{int}(q(t), in(t)) \\ f^{ext}(q(t), in(t)) \end{pmatrix}.$$

All the other relations of a system to the rest of the world, which do not directly refer to its transformational I/O-behavior, but relate to its context, like performance or security, etc., are subsumed with the notion of quality [Reich (2016b)] and are not treated in this article.

## Action, event and interaction

It is important that the notions of system, action, event and interaction fit to each other. A system is characterized by its function and the functional relation between the values of its input-, output- and internal signals. The application of a system's function is called an action. An action therefore is the transition $(i, o, p, q) \in \Delta$ happening in the context of the deterministic transition relation $\Delta \subseteq I \times O \times Q \times Q$ which defines a transition function $f: I \times Q \rightarrow O \times Q$ with $(o, q) = f(i, p)$. This is the system function of the acting system (see *Fig 4*).

In the non-deterministic case, our knowledge does not suffice to define a transition function and even spontaneous transitions occur. Here, we name a transition an "event". In a network context, only a projected part of a process in the sense of a reactive system is known during an interaction and therefore events are coupled, possibly without any identifiable super ordinated system function. This requires the distinction between describing the actions of processes (internal view) from the interactions of processes (external view).

In an interaction, the behavior or a process is characterized by a nondeterministic transition relation $\Delta \subseteq I^\varepsilon \times O^\varepsilon \times Q \times Q$, where $I^\varepsilon = I \cup \{\varepsilon\}^3$ (see *Fig 4*). Such a part is called a role. That is, a role is a process viewed from the interaction perspective. Here, a single transition can be viewed as an event, possibly triggered by some input and possibly producing some output.

The interactions of roles are described by protocols. To make sense, a protocol requires additional information like a definition of the common interaction goal and the fulfillment of the requirement of completeness with respect to further external input. Additionally, to be formally describable, an execution model is required, telling us, which transition can or has to be chosen to be taken next.

The system definition thereby naturally leads to an automaton-based definition of action, event and interaction of systems. Please note that the system definition entails a relation between the input signal at time *t* to the output signal at time *t+1*. This is in contrast to some approaches in the literature [e.g. Lunze2012, p. 94] where the output signal relates to the same point in time. This latter approach leads to substantial complexity dealing with sequential or even recursive system composition.

---

[3] $\varepsilon$ is the empty character

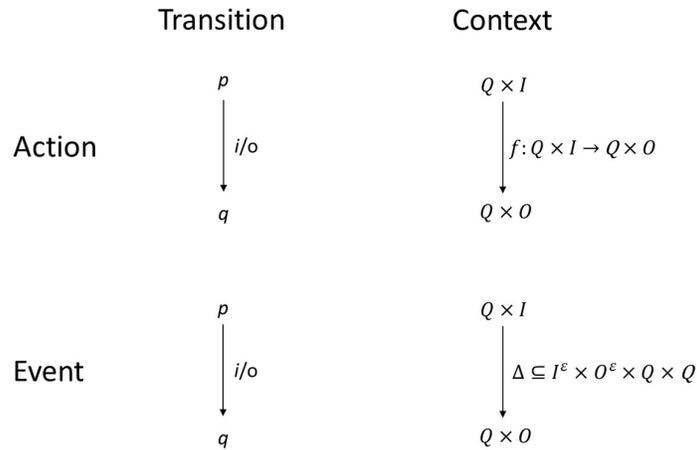

*Fig 4: The difference between an action and an event is their context of occurrence. Both indicate state changes. But an action results from the execution of a function, while an event does not necessarily have to. Hence, an event may occur spontaneously.*

Systems interact by the external coupling mechanism of information transport, that is by reproducing the value of the "sender" system's output signal onto the "receiver" system's input signal at defined times. Thus, an interaction requires us to declare two attuned transitions, where the output value[4] of the "sending" transition equals the input value of the "receiving" transition (see *Fig 5*). Thus, the coupling mechanism of interaction formally relies on identical names for the "exchanged" values, similar to a language.

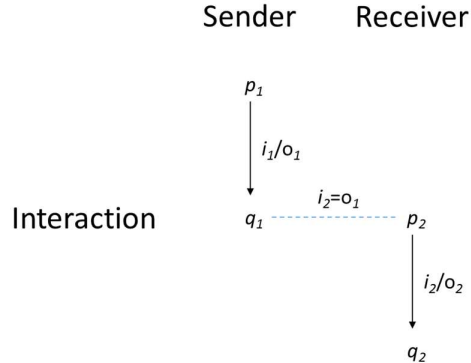

*Fig 5: An interaction implies an external coupling of two state transitions by the "exchange" of a value between a "sending" and a "receiving" transition. This coupling mechanism relies on the same name for the input and output value.*

System interaction may lead to super system formation in case a super ordinated system function can be identified on the basis of the interaction. This always happens if two actions are coupled in this way. Please note, that this understanding implies that it does not make sense to speak about interactions between systems and their eventual subsystems.

Based on the composition structure of computable functionality [Kleene1936], I distinguished the following cases of system composition [Reich2016]: sequential and parallel system composition, loop composition and while composition.

---

[4] often named "character" in the automata literature.

## Classes of interactions

According to the proposed model of interaction semantics, every sender expects its sent values to be interpreted in a given receiver context, provided by the transition relation of the receiver as far as it is known to the sender. Hence, interactions can be semantically classified according to both, the sender's and receiver's transformational behavior along the following 3 dimensions [Reich 2015]:

- Determinism of the receiver's transformational behavior with respect to the values provided by the sender.
- Statefullness of the receiver's transformational behavior with respect to the values provided by the sender.
- Synchronicity of the sender's transformational behavior with respect to the receiver.

In bidirectional interactions the roles of sender and receiver are switched. Then the following classes of (bidirectional) interactions can be distinguished:

**Symmetric:** the participants behave according to the same criteria if they are in the sender and receiver role, respectively. The most relevant combination is "mutual hinting" where all interaction parties behave nondeterministically and stateful as receiver and asynchronous as sender.

**Asymmetric:** the factors attributed to the systems as sender or receiver are different. These are the base of directed semantic relations and are therefore a possible base to define software layering. The most relevant combination is "use + observation": A user behaves synchronous as sender and nondeterministic as receiver and the used component behaves deterministic as receiver.

## Information, semantics and data

### Information

As said before, the world of information opens up, if we focus on what can be transported between state quantities which is the basic idea of Claude E. Shannon's theory of Information [Shannon1948]. Information transport means that a value represented by some state quantity at some point in time gets reproduced by some other state quantity at some future point in time. I then talk about "signals" instead of "state quantity".

### Semantics

Attributing meaning to signal values means that these values are processed, i.e. transformed. For interactions, I therefore propose a theory of local semantics, where the semantics of information cannot, per definition, be transported, but is only locally attributed by processing. Thus, within an interaction, signal values carry information but do not "have" meaning in any sense attached to them. Instead they become meaningful only by local processing, which I call "interpretation". Not-interpreted or unprocessed state values are meaningless.

The interaction semantics attributed by information processing systems rest in their transformational behavior. This notion of semantics is aligned with Leibniz semantic substitution principle as a system can be replaced by another system if it shows the same transformational behavior (and fulfills other - quality requirements).

Now I turn to the question what kind of semantics can be attributed to a piece of information, that is, a transported signal value? According to Leibniz' principle we can immediately say that two pieces of information have the same meaning if they are interpreted, that is processed, in the same way. However, "in the same way" can have many shades as I will show in the next sections.

## Data semantics

I capitalize on this tight relation between semantics and processing by the very common technique to attribute predefined meaning to (state) values by typing. With data types we declare state values to have a given structure such that they can be processed by a given set of operations. In this context of processing I will call such a state value a datum. Loosely speaking, we can say that data is typed information.

The proposed type concept is to be expressed in the language of mathematics as a pair of two sets $T = (V, F)$: a set of state – or data - values $V$ (a signal can take) and a set of (computable) operations $F$, capable of processing these values[5]. Belonging to this set of operations defines a certain "processed in the same way", which is the reason why "typing" in this sense is a semantic concept.

Please note, that this type concept does not presuppose any concept of a programming language or any other formal calculus, which is often assumed in the area of computer science, for example when Luca Cardelli says, "the fundamental purpose of a type system is to prevent the occurrence of execution errors during the running of a program." [Cardelli1997]. The main purpose of the presented type concept is to make clear, what classification of processing functions contribute to our understanding of the semantics of state values. But surely, we can use this knowledge in our design of programming languages for the important purpose Luca Cardelli points out.

This type concept also does not rely on any specific naming of the state values. It just relies on the fact that the naming convention we use to denote the state values is also followed when we talk about the operations.

Modern programming languages usually comprise some facilities to declare signals (usually called 'variables') as well as operation arguments and results to be typed, allowing the interpreter to check type affiliations. Often they also allow to declare type relations in the sense of restricted or extended sets of values or functions, but often only incomplete.

There are 4 possible data type relations. What I call "Restriction/Expansion" is useful for restricting an existing data type. What I call "Truncation/Extension" is useful for extending an existing data type. In this sense restriction is not the opposite of extension!

**Restriction/Expansion:** Be $T_1$ and $T_2$ two data types. $T_1 \, restricts \, T_2$ (or $T_2 \, expands \, T_1$) if $V_1 \subseteq V_2$ and $F_1 \supseteq F_2$. In tabular form:

| Data type relations | $F_1 \subseteq F_2$ | $F_1 \supseteq F_2$ |
|---|---|---|
| $V_1 \subseteq V_2$ | | $T_1 \, restricts \, T_2$ |
| $V_1 \supseteq V_2$ | $T_1 \, expands \, T_2$ | |

Example: Be *Char40* the type with all character sequences of length 40 as value set.  It is possible to define the restricted type *Alphanum40*, relating to all alphanumeric character sequences of length 40, by restricting the value set in relation to the value set of *Char40*.

- $V_{Alphanum40} \subseteq V_{Char40}$: Strings of length 40 with alphanumeric characters are just a subset of strings of the same length of all possible characters
- $F_{Alphanum4} \supseteq F_{Char4}$ : Each operation capable of processing all elements of $V_{Char4}$ is also able to process all elements of $V_{Alphanum4}$ .

---

[5] In contrast to abstract data types or objects in the object oriented sense, which at first sight seem to be similar combinations of sets of values and functions, a data type in the mentioned sense is a mathematical concept where especially the sets of functions, capable of processing the values, are potentially infinite. Thus, abstract data types or objects are rather defined systems with declared, fixed processing function(s) than data types according to the presented definition.

Thus, a state quantity of type Alphanum40 can be safely (expanded) casted to Char40, but not vice versa.

**Truncation/Extension:** Be $T_1$ and $T_2$ two data types. $T_1$ extends $T_2$ (or $T_2$ truncates $T_1$) if there exists a projection function $\pi: V_1 \rightarrow V_2$ such that $\pi(V_1) \subseteq V_2$ defining a type $\pi(T_1) = (V_{\pi(T_1)}, F_{\pi(T_1)})$, where $V_{\pi(T_1)} = \pi(V_1)$ and $F_{\pi(T_1)}$ is the set of operations capable of processing all elements of $\pi(V_1)$ with $F_{\pi(T_1)} \supseteq F_2$. In tabular form:

| Data type relations | $F_1 \subseteq F_{\pi(T_2)}$ | $F_{\pi(T_1)} \supseteq F_2$ |
|---|---|---|
| $V_{\pi(T_1)} \subseteq V_2$ | | $T_1$ extends $T_2$ |
| $V_1 \supseteq V_{\pi(T_2)}$ | $T_1$ truncates $T_2$ | |

Example: Be *Alphanum20* a type having all alphanumeric character sequences of length 20 as value set. Then we can construct a type *Char40* as an extension by providing a projection $\pi: V_{Char40} \rightarrow V_{Alphanum20}$. Then we have

- $V_{\pi(Char40)} \subseteq V_{Alphanum20}$: Each element in the projected set $V_{\pi(Char40)}$ is also part of the value set $V_{Alphanum20}$ of the truncated type.
- $F_{\pi(Char40)} \supseteq F_{Alphanum20}$: Each operation capable of processing all elements of *Alphanum20* is also able to process all elements of the projected set $\pi(Char40)$ of the extended type *Char40*.

With the truncation operation being the projection, a signal of type *Char40* can be safely (truncated) casted to *Alphanum20*, but not vice versa.

In summary, there are two safe casts for data types:

- Expansion cast: If a type $R$ was introduced as a restriction of an already declared Type $E$, then a state quantity $r$ of type $R$ can securely be casted to $E$.
- Truncation cast: If a type $E$ was introduced as extension of an already declared type $T$, then a state quantity $e$ of type $E$ can securely be casted to $T$ by projection.

It is possible to declare similar type and type relations for systems (or abstract data types or objects) but, due to the predefined combination of data + operations, a systematic treatment becomes more complex than for data types.

As from an engineering perspective state values are processed by operations, it is indeed possible to type every signal. But, the type concept also shows that the semantics of the data-part of a data type cannot be explicitly provided, as the set of all operations working on this data of a given type is not computable – and, practically even more important, it contains really all operations, that is, all desired as well as all undesired ones.

One can use data type hierarchies to make data semantics more concrete. Constructively, there are two ways to create data type hierarchies: either by starting from some top level type and restrict it more and more, or by starting from some bottom-level and extend it more and more. However, in both cases, safe type casting is going in the opposite direction and does not provide any new information beyond what has been constructively put into it, as is shown in *Fig 6*.

# Two ways to create a type hierarchy

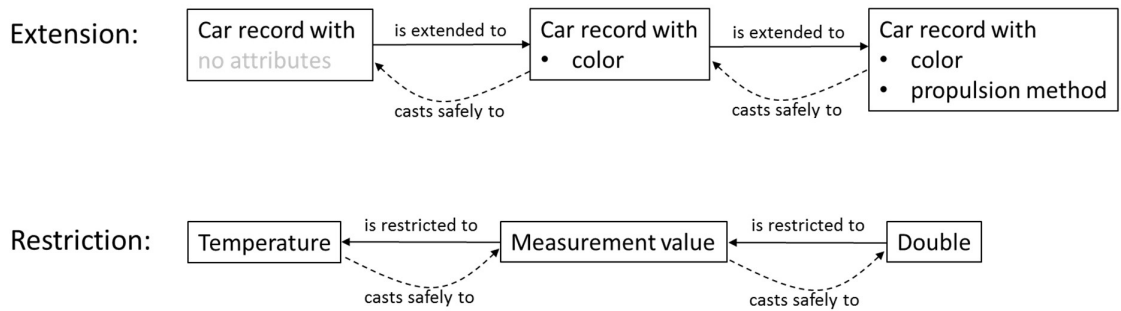

*Fig 6: Two ways to constructively create a type hierarchy. Neither safe type casting results in any new knowledge.*

So, it is not type casting by which we get a more concrete meaning, but the construction of the type hierarchy beforehand, which has to be guided by some sort of understanding of the to-be-represented semantics. This understanding is provided by the model of open semantics of open interaction networks.

## The semantics of interaction networks

Interaction networks in the sense of this article are characterized by their openness and by the locally limited sender knowledge of the receiver's processing context. What are the implications for their semantics?

If we say that two pieces of information have the same meaning if they (that is the state values) are processed in the same way, then this is the case if both pieces of information received with the same internal source state value lead to the same internal target state value together with the same output value.

However, in contrast to the deterministic case, in a nondeterministic interaction the internal state value is possibly only partially known to the sender both, before and after the interaction. Additionally, it could well be that some of the information is forwarded to a third party, which process it further and even somehow interacts with the first party in the context, the first party issued the information beforehand. For example: A patient goes to a hospital in a treatment context. There she has to provide her insurance number. For treatment, this number is irrelevant. The hospital needs it from the patient only because it wants to bill its services to the patient's health insurance. Depending on the contract between insurance and patient, such a visit may unfold some relevance for the patient-insurance relation, as some insurance reimbursement may depend on the patient-hospital interaction. We see that in a network context we cannot deduce all relevant semantics of a piece of information from a single interaction, or, in other words, the semantics is open.

As non-deterministic transition relations cannot be described on the element level by functions, the semantics of the exchanged data (elements) cannot be fully conceived with data types in the proposed form, as they only relate to functions between elements.

## The I-language

It's important to note that from an engineering perspective we are talking at least about three different languages when dealing with interactions: there are languages we use to talk about the interaction. These are our normal (engineering) language and our formal programming languages. And there is the interaction itself which can also be viewed as a language. To distinguish it from the other two Felix Bangemann, Johannes Reich and Christian Diedrich [BaReDi2016] have introduced the term "I-language". Already Gerard Holzmann [Holzmann1994] compared protocols with languages. The late Ludwig Wittgenstein explained the semantics of natural languages by game-semantics, which is very similar to protocols [Reich2009]. So, the problem of understanding the semantics of interaction networks is equivalently formulated as the problem of understanding the semantics of the I-language.

The semantics of formal languages is usually strictly compositional, as the meaning of the composite-entities can be deduced entirely from the meaning of the parts down to some predefined semantic "atoms". This is neither the case for our normal language nor for the I-language, because of their context dependencies – which suggests that our normal language can also be viewed as an I-language. In my opinion, natural language can be viewed as the solution to the problem that having a non-deceptive interaction, free of misconceptions and misunderstandings, always requires the establishment of a common context of interpretation beforehand. As this context cannot be ascertained adamantly, it is always a question of a situational consideration whether it is still possible to proceed with the original discourse or whether it is advisable to affirm the context – requiring yet another context. That is the deeper reason for our astonishing capability to switch this context even within single sentences.

But even within technical interaction it is already common, to talk within an interaction over the interaction itself, e.g. as negotiation about security levels or, more generally, about how to proceed.

The important question remains, which parts of the interaction semantics can we define by a type system and which parts have to remain in the concrete transition relations of the interactions?

The proposed type concept tries to express that our common knowledge about the state quantities reaches as far as we know how to deal, that is process, their values. For example: if we talk about a temperature, we might not know what kind of temperature it is, whether it is the temperature of an automobile engine or a room temperature, but we suppose that in case we would know all these aspects of additional set up we could sensibly deal with it. If we move our temperature sensor on our car engine some centimeter to the left or right, it still measures "temperature", but the practical relevance of this other temperature might be completely different in its interaction context, that is, it has to be processed differently but at the same time similar, namely still as a temperature. We could derive a more restricted temperature type relating only to temperatures on car engines of a given type at a given position – or we could decide to use the generic temperature type and keep our specialization knowledge in the area of the concrete transition relation.

Hence, it seems that it is an important design decision how much differentiating knowledge we put into our type system and how much we leave in the concreteness of the specific interactions.

### Finding equivalence classes

Based on Leibniz substitution principle, finding useful equivalence classifications are a fundamental tool for analyzing network semantics, answering the question: which values are attributed similar semantic aspects by which kind of processing? An equivalence classification provides a disjoint partition $\{\Delta_l\}_{l \in L}$ of the transition relation $\Delta = \cup \Delta_l$ of a processing system such that according to a given criterion (the equivalence relation) every value can be attributed uniquely to exactly one subset, the equivalence class $\Delta_i$, that is $\Delta_i \cap \Delta_j = \{\}$ for $i \neq j$. Data types of the same level of a type hierarchy can be understood as such an equivalence classification.

In the following, two additional equivalence class constructions, one for the deterministic and one for the nondeterministic case, are introduced.

**Deterministic transition relations:** any equivalence partition of a deterministic I/O-transition relation results in disjoint deterministic I/O-subrelations, each defining a transition function $(o, q) = \delta_l(i, p)$ on their own. Combining the external part of these equations with data typing we arrive at the current object oriented way of describing state transitions of "objects" (or abstract data types) whose syntax is in many object oriented programming languages more or less outputParameters = objectName.method(inputParameters).

Partitioning the internal state into a mode and a rest component $Q = Q_{mode} \times Q_{rest}$, we create the possibility to define generic events which could be used for a generic bottom-up interaction in a leveled architecture $(o, q) = \delta_l(i, p) = \delta_l(i, p_{mode}, p_{rest}) = \delta_l|_{p_{mode}}(i, p_{rest})$. The mode component can thereby be viewed as modulating the transition function.

Such a separation of the internal state, one part for modulation of the behavior, the other part for parameterization of the behavior itself, is called "state pattern" in the object oriented community (Gamma et al. 1995). In this sense, a change of the mode value signals a change in the behavior of the object.

**Nondeterministic transition relations:** A natural partition of the nondeterministic transition relation leads to what I call "extended I/O-automata".

We create the partition function $partition: I^\varepsilon \times O^\varepsilon \times Q \times Q \to \mathbb{N}$ with the following assumptions:

1. For all I/O-characters $a \in I \cup O$ there exists a parse-function $parse: I \cup O \to DocCls \times Param$, with $(docCls, param) = parse(a)$ which attributes each character $a$ a document class and values of additional parameters. Thus we can interpret differently parameterized documents of a document classes as different characters of I/O-alphabets (or characters as parameterized documents of a document class).
2. The internal state value is partitioned in a mode and rest part: $Q = Q_{mode} \times Q_{rest}$
3. There is a set of conditions operating on the internal rest state and the parameters of the incoming documents. $Cond: Q_{rest} \times Param \to \{true, false\}$

Then we can construct a partition function as

$$partition(i, o, p, q) = partition(docCls_i, docCls_o, p_{mode}, q_{mode}, cond(p_{rest}, param_i)).$$

$p_{rest}$ und $param_o$ remain unconsidered in this partitioning. Thereby we can write for each equivalence class:

$$\Delta_l = p_{mode} \xrightarrow{docCls_i, cond(p_{rest}, param_i)/docCls_o} q_{mode}$$

A sensible requirement is to attribute all characters which result in the same class of internal mode state and lead to characters of the same class of (outgoing) documents into the same document class – thereby expressing their similar meaning according to our definition. Thus, the equivalence class construction is not completely arbitrary but is naturally guided by the structure of the transition relation itself.

**Exceptions:** No rule without exception! If, for whatever reason, we want to stick to our deterministic point of view even in case of a nondeterministic transition relation, we can partition this relation into a deterministic part, which defines a transition function and an exceptional part, comprising the rest. Most modern programming languages provide syntactical means for handling desired – deterministic

– versus exceptional – nondeterministic – behavior, for example by try-catch-clauses where the deterministic part is described as an operation and the exceptional behavior as an event mechanism, modifying the control flow of the calling component.

## Software reference architecture

The IoT software reference architecture provides the structures and respective elements and relations for concrete software system architectures in the IoT domain. We now have to use the insights from the reference model of interaction semantics and ask ourselves what kind of structures and respective elements of software systems can be derived or should be demanded.

The following software reference architecture provides

- A definition of software layers
- Means to express vertical interactions, that is interactions which demarcate a software layer.
- Means to express horizontal interactions, that is, interactions between processes in the same software layer.
- A definition of a component and how to distinguish it from other entities like systems or objects.
- A model how to separate reusable from non-reusable parts of an application's functionality.

A major guidance is the requirement to minimize the effort to adapt software systems to changes in their interaction semantics. One obvious consequence is the requirement that a software system should be created from components: building blocks which easily fit together. As Clemens Szyperski expressed it: "Components are for composition" [Szyperski1998]. A very important, although not as obvious additional implication based on the reference model then is, that a component has to be qualified by both, its I/O-relation from an interaction perspective and its intended composition behavior. This consideration leads to several distinct classes of component interfaces for unidirectional and bidirectional interactions with different formal, that is, syntactical representations.

### Classes of bidirectional interfaces

**Operations:** Operations [or functional API] appear in the programming code of imperative programs representing the "called functionality". Thus, operations semantically represent mathematical functions with a special composition behavior. After successful computation of the operation, the control flow picks the next operation in the program. From a compositional perspective the control flow hand over to the called operation implies a sequential system composition and the return of the control flow to the calling operation implies a coalescence of both systems into a trivial recursive loop system with one iteration step.

Remote operations take advantage of the fact that the functions of serialization and deserialization of data, transport, and data processing can be concatenated into one function which can then be represented by a "remote" operation. An important difference between remote and local operations is the nondeterminism introduced by the imponderabilities of the data transport. Thus, remote operations usually come along with additional so called "remote exceptions". So, a remote operation is a special case where it might be sensible to keep the fiction of determinism even in case of nondeterministic behavior by separating the desired "normal" deterministic from the other "exceptional" behavior. But otherwise there is no further semantic difference.

For concatenated operations comprising complex state changes, their transactional control might be important to achieve an all –or-nothing execution semantics to stay consistent. Due to principal reasons (see the byzantine generals' problem) this cannot be achieved for remote operations as sufficiently as for local operations. Hence, changing some complex remote state by a remote operation is often not a good idea.

Functionality can be constructed hierarchically, creating a hierarchy of functionality. To provide all computable functionality, also recursive loop and while-constructs are necessary [Kleene1936]. However, as there are no simple interfaces in the sense of an operation for these constructs, they are usually handled by control-flow statements within operations.

As an operation semantically only relates to the transformational I/O-behavior, it still provides the freedom to choose the implemented algorithm.

Operations provide a natural mean to express hierarchical component relations.

**Objects [or abstract data types]:** Objects [or abstract data types[6]] are data, encapsulated with a finite set of operations and a special composition behavior. The encapsulation fixes the processing of the internal data and thereby represents a much stronger semantic stipulation for this data than typing.

This encapsulation is necessary for

- (even further restricted by the state pattern) the definition of generic events (Generic event generating object: GEGO);
- creating an automatic reference of generic event-consumer context, as every data context which references an object capable of generic eventing must provide an appropriate event handler.
- enabling reuse of stateful processing entities.

However, its positioning against typing, which represents a much looser coupling between data and its processing, for application architecture is not entirely clear.

From a composition perspective, an object with reference to several other dependent objects has correspondingly many bidirectional interactions. Each such interaction leads to a combined sequential plus trivial (one step) loop composition. Thus, from a system composition perspective, the called object is not a subsystem of the calling object. Instead, while calling the dependent object, the called and the calling object are both sequentially coupled subsystems and with returning the processed data into the calling objects, both systems coalesce into a single loop system.

**Protocol roles:** due to the nondeterministic relation between the relevant state values, we refrain from explicitly describing what the other system is supposed to do in imperative code[7]. To interact protocol-based, documents are exchanged that signal the state transition of the sender in a way that a receiver can appropriately process them ("understand them") to make a sensible transition by itself. Generally, no operations of the interacting partners beside those representing transport functionality are used.

A protocol does leave real decision freedom and is very similar to a game. Also very alike games, protocols can be partitioned into roles, where each role describes the possible behavior of a single participant. It can naturally be used to express non-hierarchical component relations.

---

[6] I do not consider details like polymorphism here, where objects in the object oriented sense might differ from abstract data types as they are usually understood.

[7] The alternative would be to use functions which map values to sets of values. However, this approach leads to exponential complexity with the number of possible alternatives in each interaction step.

## Classes of unidirectional interfaces

**Pipes**

A pipe is a strict sequential processing of data. A pipe mechanism provides the means to couple the output to the input of systems without providing any means for a "backward" channel. From a compositional perspective, pipes thereby provide the means for strict sequential and parallel system coupling. To be complete, a pipe mechanism must be able to fork and to join pipes.

With adapters that changes the composition behavior, pipes can be transformed into operations and vice versa.

**Generic events**

Generic events require "generic events generating objects, or in short "gegos". These are objects with an implemented state pattern, where the internal state is partitioned into a mode and a rest. The mode part determines the set of active operations and thereby the sort of behavior while the rest part is the state that the operations operate on. With this state partition an explicit behavior model of the object results which can be represented in a standardized format and which can be related to by a generic event subscription mechanism.

The question, which part of the object behavior is put into the explicit behavior model and which part is put into the rest, is a design decision and depends especially on the expected context of usage and how detailed the potential users want to be informed about behavioral changes of the dependent object.

So, although strictly speaking, generic events are a mean for unidirectional interaction, they make only sense in the context of an already directed bidirectional interaction.

The subscriptions to a generic event is not part of the semantic guarantees that determine the correctness of the gegos operation implementations. Thus for a gego, it does not make any semantic difference, whether zero, one or many users subscribe to its state change indications.

## Software layers

A statement "component A resides in software layer XYZ" has to be provable. It's not enough that it can be asserted as the OSI-model did, for example by putting a nice little box in a layered diagram, but this property has to be deducible from the composition behavior of the components itself.

First we note that the system notion is not appropriate for comprising the layer-semantics of software applications. There simply is no interaction between a system and its subsystems, but systems (in the sense of a super system) come into existence because of the interactions of their subsystems. In contrast, components of different software layers do interact.

Second we note, that simple unidirectional interaction, although it defines an interaction direction, makes any classification into "horizontal" versus "vertical" an arbitrary convention. It is the possibility of bidirectional interactions which enables the very important distinction between "horizontal" and "vertical" – or "directed" versus "undirected".

The definition of software layers can be based on the classification of interactions between software components according to the dimensions mentioned in the reference model section: determinism, synchrony and statefullness.

# Classes of bidirectional interaction

**Vertical = Use + Observation**          **Horizontal = Mutual Hinting**

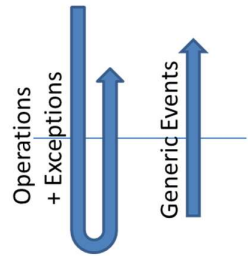
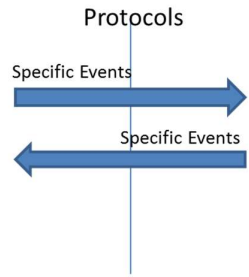

- Asymmetric: Component − Subcomponent
- Function-oriented
- Top→Down: Deterministic, synchronous
- Bottom→Up: Nondeterministic, asynchronous
- Tight coupling
- Used for system composition
- E.g. object model

- Symmetric: System − System
- Document-oriented
- Both sides are
    - Stateful
    - Asynchronous
    - Nondeterministic
- Loose coupling
- E.g. Negotiations, Games, …

*Fig 7: The two most important classes of interactions between software components*

**Hierarchical interactions: use and observation**

The asymmetric interaction differs in its directional components: The top-down interaction is deterministic and synchronous and is represented by operations, that is, traditional API-calls. If there is no call back from the called context, then a traditional API-call is a formal indicator for stepping into a lower semantic layer.

The bottom-up interaction is characterized by an event semantic, enabling the semantically superior entities to observe the behavior of the semantically subordinated entities.

To provide a complete interaction model for layered software components, a component runtime environment has to provide both facilities, top-down operations as well as bottom-up events. If the component context does not provide a bottom-up eventing mechanism, then explicit API-calls become necessary to signal state changes "upwards", destroying the meaning of API-calls as a formal criterion for layering and thereby destroying the provability of any layered structure of an application.

**Non-hierarchical interactions: mutual hinting**

In the most important case of symmetric interactions all interaction parties behave non-deterministically and stateful as receiver and asynchronous as sender[8]. These kind of interactions are described by protocols.

---

[8] The other case, where all parties behave deterministically, only works for clocked systems. But more importantly, because of the resulting recursive functionality, such kind of interactions should absolutely be avoided on a component level, as it leads to recursive functionality and thereby quickly becomes too complex to be handled on this level of granularity.

With protocols, there are no imperative statements any longer in the application about the behavior of the interaction partners. Instead, documents are created which document the state change of the sending applications in a way that a receiving application can sensibly process it.

With their roles, protocols also provide an effective way to define precisely what kind of knowledge remains private and what kind has to be shared to cooperatively interact to achieve a defined goal[9].

## Components

Components are viewed as building blocks which easily fit together. They can be classified by the qualities of their I/O-relation and their intended composition behavior as is illustrated in *Fig 8*.

Recursive interactions in a loop or while sense, creating complex recursive functionality does not follow any simple composition schema and should be avoided on the level of components. Components thereby mark a systematic border of complexity, where any functionality that is created by recursion moves into the component's innards to enable the desired simple composition behavior.

Hierarchical composing components provide "consumable" functionality together with generic events. Hence, their interface is to be described by APIs and the behavior model of gegos.

Non-hierarchically composing components integrate in interaction networks. Hence, their interface is to be described by roles of protocols.

A third composition class, representing unidirectional interacting components, are pipes.

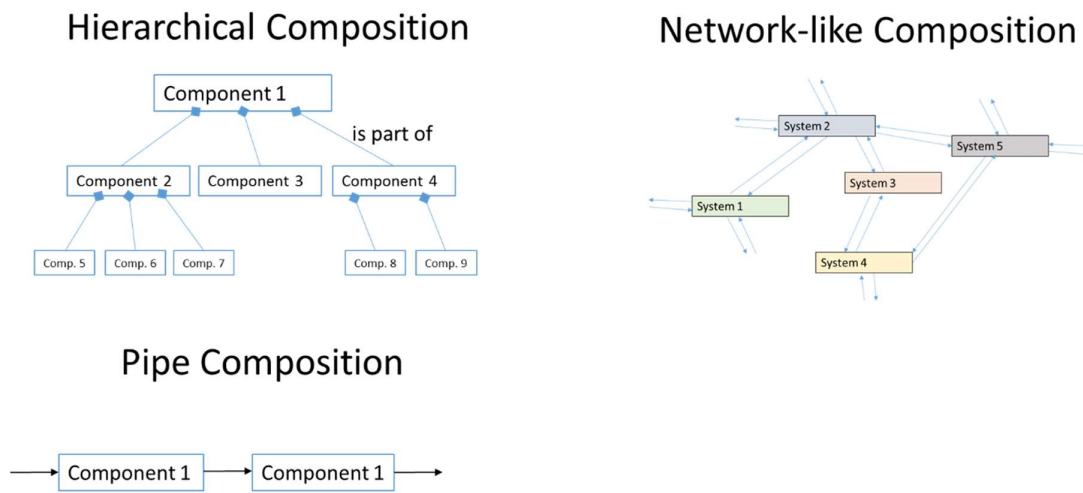

*Fig 8 the three main composition classes for components.*

## Reuse

As explained in the reference model section, for processes we get an ostensible dichotomy: on the one hand, processes – like any other systems - are described by their functions, on the other hand, their interactions are best described by protocol roles.

## Vertical reuse

One of the major parts of any application architecture model is to separate reusable from non-reusable parts. In a layered architecture, that is in a software structure without circular API-references, this is achieved by API-definitions. With APIs, functionality can be encapsulated for reuse in different contexts.

---

[9] It therefore would make much sense to provide legal regulations that the specifications of protocols of public interest have to be public knowledge, as it has to be shared knowledge anyway.

The object model of an application can be understood as the combination from state and functionality that can be reused in different process contexts and remains either internal/private or partly shared with other systems.

I name this "vertical" reuse. It also implies where all non-reusable elements have to be collected: at the semantic top layer, which I call "process layer". A process layer therefore provides an architect a dedicated space to organize the non-reusable functional elements.

### Horizontal reuse

However, the interaction oriented architecture goes an important step further and provides an additional modus of reuse, namely the reuse of role implementations, which I call "horizontal reuse". A reuse of roles implies that all interaction guarantees, like completeness of the protocols as well as their freedom of deadlocks or livelocks are preserved, despite a changed role coordination within a process. A key issue of application architecture therefore becomes the question on how to synthesize processes from role-implementation. This question can be reformulated as to find the internal coordination between the different events of a process. This can be achieved by a set of rules, relating the event of the different roles of a single process together according to the internal perspective. Rules, determining the sending of documents to signal a state change of a process, are, by definition, part of the semantic top (process) layer – and cannot be delegated to any semantically lower "middleware".

## Relation to other models and architectures

### Open Systems Interconnection basic reference model

The Open Systems Interconnection (OSI) basic reference model [OSI1994] assigned subsystems to a given layer by the criterion that they are supposed to interact only with the next higher or the next lower layer in the same system. A layer was then (circularly) defined as all subsystems of the same rank. The active elements of a subsystems were referred to as "entity".

The term "service" was used to denote the capabilities of a layer (with the eventual help of entities of lower layers) that was provided to the next higher layer. The term "protocol" was used to denote a set of rules and formats which determines the communication behavior of the (entities of the) subsystems in the performance of their functions.

Although the ISO/OSI-model did refer to interactions to qualify horizontal versus vertical component relations, it was not sufficiently specific about them. As the layering criterion, the ISO/OSI model did refer to the "use"-relation and also distinguished between request/response for top-down and indication for bottom-up interactions but it did not explicitly defined what "use" concretely meant or how a "response" could be formally distinguished from an "indication". Also there was no clear system definition which would have allowed to deduce the system-subsystem relation.

As the remote operation technology showed, it is not communication per se which decides about which entities in which layer interact in different applications. So, it is not a priori clear – as the OSI model assumed – that a protocol enables access only to the same layer of another interacting application. As the remote operation technology, which is implemented by protocols, shows nicely, one can also access lower layers.

One could view the concepts of the proposed interaction oriented architecture as an advancement on the ideas of the ISO/OSI-model, complemented by sound semantic concepts.

## Service Oriented Architecture (SOA)

Within the OASIS SOA RM, a service is defined as a "Mechanism to enable access to one or more capabilities where the access is provided by a prescribed interface and is exercised consistent with constraints and policies as specified by the service description."

However, at no place within SOA the service definition refers to the transformational behavior of a service – with dramatic consequences.

The WSDL 1.1 Specification defines 4 "transmission primitives" which a so called "endpoint" is supposed to support: one-way, request-response, solicit-response and notification. WSDL denoted these "transmission primitives" as "operations". In its section 2.2.1 WSDL2.0 defines an "Interface component" as a "sequences of messages that a service sends and/or receives". According to WSDL 2.0 an "operation" is an "interaction with the service consisting of a set of (ordinary and fault) messages exchanged between the service and the other parties involved in the interaction". BPMN-Version 2.0.2 formally relates to the WSDL 2.0 specification (although only non-normatively). In section 8.5.3 it states that "An Operation defines Messages that are consumed and, optionally, produced when the Operation is called."

Thus, from the point of view of this reference model, the semantics of a service in the SOA-sense was not well defined. However, from a syntactic point of view, SOA ties oneself down to (remote) objects with methods. Thus, well defined services specified by a WSDL-specification can only represent accessible functionality. This is probably the main reason why SOA did not succeed as desired in the area of loosely coupled applications although it was backed by virtually all major software companies worldwide at that time.

Also, service "choreographies", the analog to protocols, suffered from the inherent contradiction that using WSDL, protocols in the sense of this IOA cannot be described adequately. Additionally, the bipartition of the interactors in service "provider" and "consumer", possibly borrowed from the hierarchical model of interactions, is inadequate to describe the interaction of networking systems in their different roles. Often, as in games, there are not just two but multiple parties participating in an interaction.

So, by naming, SOA suggested to provide some IT-equivalent to modern economic service in the sense of the German "Dienstleistung". But in reality it just provided a further mechanism for operating system independent remote function calls – and not a mechanism to describe "services" in an economic sense. Why that? "Services" in an economic sense are usually highly interactive. For example, to get a house painted on a construction site by a painter one does not command "paintHouse(money)", but proceeds as following: an offering has to be specified, the offering of several painters has to be compared, one has to be accepted, the dates where the craftsman work on the construction site have to be agreed upon, maybe rescheduled several times because of dependencies to other trades, the work has to be checked and finally paid, perhaps in several tranches.

In retrospect, it seems astonishing that such an endeavor could more or less completely ignore the question of interaction semantics, without being recognized. As a result of its inherent contradictions, SOA possibly even hindered the growth of the electronic interaction networks in the last couple of years.

## Representational State Transfer (REST)

REST [see Fielding2000] can be viewed as the attempt to transfer the principles of stateless communication together with semantic agnostics - both principles of the HTTP-protocol - onto the interactions of networking applications. Currently it is often positioned as a simpler variant of SOA.

A REST-service is supposed to adhere to at least the following three principles:

- Addressability: each resource has to have a unique URI.
- Statelessness: each REST-message is supposed to contain all the information that is necessary for the processing which it initiates.
- Idem potency: the called transport methods are supposed to have an identical effect, no matter when they are called – as the HTTP specification requires for the HTTP-operations GET, HEAD, PUT und DELETE.

These "principles" are in direct contradiction to the proposition of the proposed model of interaction semantics that loosely coupled interactions as they are describable by protocols are stateful and the exchanged data is usually not processed in an idempotent way.

From the perspective of the proposed model of interaction semantics, REST is a methodological chimera with parts from the object as well as the protocol world. On the one hand side, it only specifies a letter-box mechanism, where some of the letter lifecycle functionality was given to the sender and it poses no requirements to specify the transformation behavior of any REST-service or even the relation between different REST-services. But on the other hand side it requires that all resources have to be published to the public.

It is to be expected that application based on this "architecture style" will run into trouble in combination with the micro service approach (see below), as on the one hand it becomes possible to interact in a protocol sense – but, on the other hand, without the possibility to comprehend the complex interaction semantics, as these are more or less ignored. So, all the difficult problems of parallel processing like occurrence of livelocks or deadlocks remain inaccessible for analysis.

## Micro Services

Martin Fowler started his talk at SAP about micro services with the saying, "micro services are neither micro nor services". Micro services can perhaps best be characterized as a sensible combination between a process-based runtime model and a deployment model.

Practically, software engineers, developing micro services based on REST-APIs, very quickly discover, that from an interaction point of view their entities are neither stateless nor idempotent, confirming the process model of the proposed reference model.

Actually, the proposed interaction oriented architecture provides the necessary structure and understanding to make micro services a success, namely the necessary process definition based on the distinction between interactions and actions, i.e. the distinction between the internal and the external perspective.

A micro service landscape is to a large extend a networking application landscape with all the issues of parallel processing. To keep such a landscape running while it evolves continuously, it is absolutely essential to be able to keep any changes in single interaction between these kinds of applications to a minimum throughout the whole landscape, which is one of the strongholds of the proposed interaction oriented architecture.

## DIN SPEC 16593 Reference Model Service Architecture (RM-SAI40)

The starting point of the DIN SPEC 16593 RM-SA [RM-SAI40] was the stated requirement of the DIN SPEC 91345 Reference Architecture Model Industrie 4.0 (RAMI4.0) [RAMI40] that the architectural style of the industry 4.0 stack should be "service-oriented" and the insight that, despite more than 20 years of the so called "Service Oriented Architecture (SOA)" [ScNa1996], it is still unclear what is meant by a service in a SOA context. As one of its major goals it explicitly states to clarify what the concept "service" and "protocol" stands for.

The RM-SAI40 refers explicitly to the three object worlds as defined in DIN SPEC 91345 RAMI4.0: the human, the information and the physical world but declares that "the human world as well as the human interface to the information and physical world is outside the scope of this DIN SPEC".

However, the presented concepts are not clear. The authors renounce a formal definition of a system, but refer to the ISO/IEC 10746-2:1996 definition saying that "a system is something of interest as a whole or as comprised of parts [subsystems]".

An intra-, extra- and inter-component views are distinguished and serve as context for the definition of the concepts.

A "state" of a component C at a given time $t_x$ is defined as a set of state variables (state (C, $t_x$) = ($v_1$,...,$v_n$)). An "action" is defined as a transition from one state S1=(C1, t1) to another state S2 = (C, t2) and is generally modelled as a 4-tuple (S1, S2, in, out) with two internal state values and an input and output state value. It is said to possibly have an effect in one of the three worlds. It is triggered by some "trigger" with additional dependency on a "condition". However the difference between the trigger and the input does not become clear. Also, the entity of action + condition + trigger, although seemingly of fundamental importance is not named or discussed further.

An "operation" is said to be a projection upon an action of a component which may be triggered through an interface of a component with the restriction that not all actions may be triggered as operations, and not all state variables (input and output) related to an action may be part of the operation. An "event" is defined similarly as a projection of an action – but it may be observed (instead of triggered) by an interface. No reference to the mathematical concept of a function is made.

To be considerable as an "exchangeable format" (inter component view), operations and events are said to encoded into data packages (remains undefined), which may be either "primitives" (for operations) or "notifications" (for events). Hence, a primitive is supposed to signal either an operation request, an operation result or an operation error while a notification is supposed to signal an event.

An "interaction" is then finally defined as a "sequence of operations encoded in primitives and/or events encoded in notifications exchanged between two or more system components". So, the terms "operation", "primitive", and "event" seem to be more basic than the term "interaction" as the former are used to define the latter. According to my understanding, this construction obscures the fact, that the authors therewith define an interaction simply as an exchange of information – no more, no less.

Typical sequence of primitives and/or notifications are called "interaction patterns". The set of rules determining the exchange of notifications is called an "interaction policy".

A service is broadly defined as "a bracket around one or more interactions comprising (operation) primitives and/or (event) notifications.  A service is defined via an interaction pattern and how the execution of the pattern changes the state of the information world (that includes the service context) and the physical world in the scope of the service participants. A service is advertised and accessible by means of service interfaces."  Participating ín a service, the participants are supposed to play either the role of the service provider or service consumer.

In my opinion, with this service definition, a service can virtually be anything – making the restriction of the participants' roles even more artificial.

The definition of a "protocol" was supposed to complement the service definition as "the specification of the rules that determine the exchange of (event) notifications between interacting components." However, this is more or less what has been already defined for the interaction policy. And it is not clear how to delineate such a protocol from a service in the broad sense of this DIN SPEC – which was the original endeavor.

In summary, this DIN SPEC did start with an accurate diagnosis, but it remained too strongly tied to the "service oriented" paradigm and did pay too little attention to the intricacies of the notion of an interaction context. As a result, its proposed conceptualization seems to be unnecessarily overloaded (see the definition of an "interaction") and with increasing abstraction level accumulative unclear (see the service definition). Thereby it misses its stated aim to clarify what the concept "service" and "protocol" stands for.

# Revision History

2017-01-29 First version

2017-05-16 Correction of some minor typos and rewrite of the data type section on extension/truncation due to a conceptual mistake.